# Cherenkov light Extrapolation at Ultra High Energy Cosmic Rays in Extensive Air Showers

**Al-Rubaiee A. A.**
Al-Mustansiriyah University, College of science, Dept. of physics, Iraq

**ABSTRACT**

The Simulation of Cherenkov light lateral distribution function (LDF) from particles of Extensive Air Showers (EAS) with ultra high energy cosmic rays ($E \geq 10^{16} eV$) was simulated for primary protons by the computer code CORSIKA. The parameterization, that constructed on the basis of this simulation have allowed us to reconstruct the events, that is, to reconstruct the type and energy of the particle that generated EAS from signal amplitudes of Cherenkov light registered with the Tunka-25 facility. The extrapolation of the Cherenkov light LDF approximation at the energy range ($10^{16} - 2 \cdot 10^{18} eV$) was taken into account.

## INTRODUCTION

The determination of the energy spectrum and mass composition of the ultrahigh energy cosmic rays (E>$10^{16}$ eV) is one of the greatest challenges in cosmic ray measurements. Using the atmosphere as a large target, Detectors are capable of tracing the development of the size of the Extensive Air Shower (EAS) through the atmosphere [1, 2]. Large-scale experiments, like the Yakutsk EAS array [3], AGASA [4] HiRes [5], Pierre Auger Observatory [6] and Tunka-133 [7] focus on the precise determination of the energy spectrum, mass composition and arrival direction distribution of ultrahigh-energy cosmic rays. The analysis of the characteristics of the detected longitudinal profiles is currently the most reliable way for extracting some information about the primary cosmic ray mass composition. One of the main techniques for observing EAS is effectively investigated by the method of Cherenkov light EAS registration [8, 9]. The main tools for calculating of EAS characteristics and experimental data analyzing (Direction of the shower axis, determination of the primary particle energy and type from the characteristics of Cherenkov radiation of secondary charged particles) are codes of numerical simulation by the Monte Carlo method. Reconstruction of the primary particle characteristics initiating the atmospheric cascade from Cherenkov radiation of secondary particles at the energy ranges ($10^{16}$-$10^{18}$) eV calls for the creation of a library of shower patterns, when this requires much computation time.

In the present work, the CORSIKA software package [10] in which hadron interactions are simulated using the QGSJET [11] and GHEISHA codes [12] simulates lateral distributions of Cherenkov light emitted by atmospheric cascades initiated by primary high-energy cosmic ray protons and nuclei. Simulation of Cherenkov radiation using the CORSIKA code requires very long computation time for a single shower with energy of $10^{18}$ eV for a processor with a frequency of a few GHz. Therefore, the development of fast modeling algorithms and the search for approximations of the results of numerical modeling are important practical problems.

Parameterization of the lateral distribution function (LDF) of Cherenkov radiation versus the distance $R$ from the EAS axis and the primary particle energy $E$ that can be used to approximate the results of numerical simulation of LDFs of Cherenkov photons emitted by EAS initiated in the Earth's atmosphere by the cosmic ray particle having a very high energy was taken. In the present work, we use this parameterization to describe results of numerical simulation of EAS by the CORSIKA code and of Cherenkov light emitted by EAS measured with the Tunka-25 facility [13].

## THE CHERENKOV LIGHT LDF APPROXIMATION

The simulation of the Cherenkov light LDF in EAS is obtained using CORSIKA code for primary protons at the highest energies ($E \geq 10^{16}$) eV. For parameterization of simulated Cherenkov light LDF, we used the proposed function as a function of the distance $R$, from the shower axis and the energy $E_o$ of the initial primary particle, which depends on four parameters $a$, $\gamma$, $\sigma$ and $r_o$ [14, 15]:

$$Q(E,R) = \frac{C\sigma e^a \exp(R/\gamma + (R-r_0)/\gamma + (R/\gamma)^2 + (R-r_0)^2/\gamma^2)}{\gamma[(R/\gamma)^2 + (R-r_0)^2/\gamma^2 + R\sigma^2/\gamma]} m^{-2}, \quad (1)$$

where $C=10^3$ m$^{-1}$; $R$ is the distance from the shower axis; $a$, $\gamma$, $\sigma$ and $r_o$ are parameters of Cherenkov light LDF. The energy dependences of the parameters, $\gamma$, $\sigma$ and $r_o$ is shown on Fig.1 for primary proton.

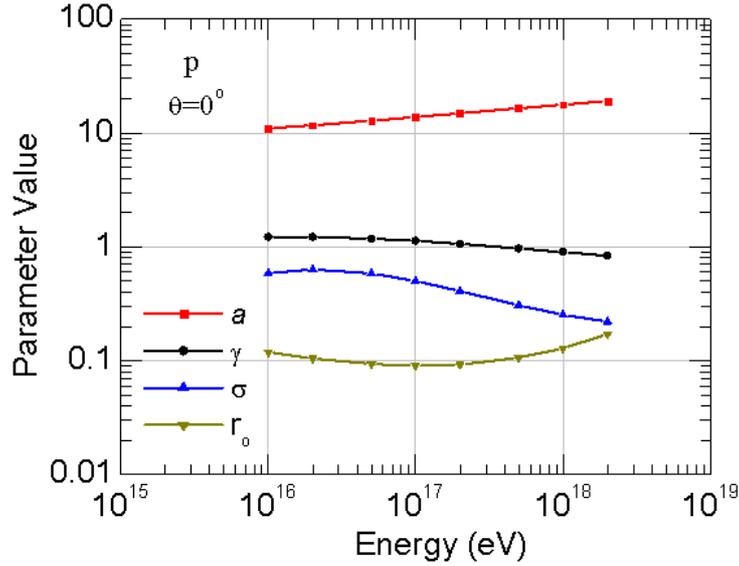

**Figure 1.** The fit parameters $a$, $\gamma$, $\sigma$ and $r_o$ as a function of the primary energy of the initiating primary proton for vertical EAS.

The simulated data and the approximated formula (Eq.(1)) for vertical showers are presented on Fig. 2. for primary protons at the energies $10^{16}$, $2\cdot10^{16}$ and $5\cdot10^{16}$ eV. In Fig. 3. one can see the extrapolation of the Cherenkov light LDF parameterization of the obtained data with CORSIKA program at the energies $10^{17}$, $2\cdot10^{17}$, $5\cdot10^{17}$, $10^{18}$ and $2\cdot10^{18}$ eV. The accuracy of the Cherenkov light LDF approximation for vertical showers for primary protons is better than 15 % at the distances 80-120m from the shower axis, and close to 5 % for the other distances.

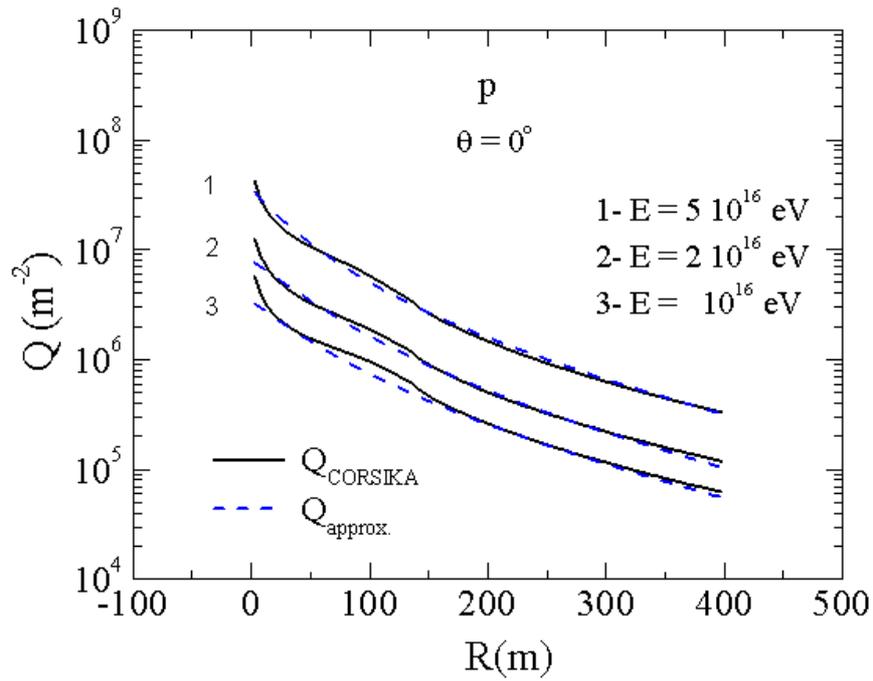

**Figure 2.** Lateral distribution of Cherenkov light which simulated with CORSIKA code (solid lines) and one calculated (Eq. (1)) (dashed lines) for vertical showers initiated by primary protons at $10^{16}$, $2 \cdot 10^{16}$ and $5 \cdot 10^{16}$ eV.

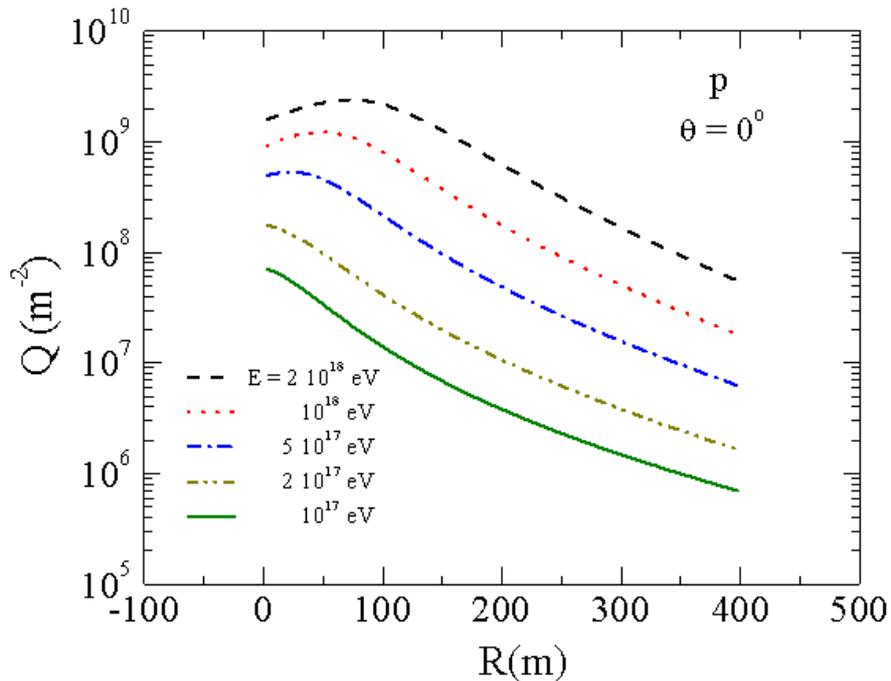

**Figure 3.** Extrapolation of the Cherenkov light LDF parameterization at energies $10^{17}$, $2 \cdot 10^{17}$, $5 \cdot 10^{17}$, $10^{18}$ and $2 \cdot 10^{18}$ eV with the help of Eq.(1).

# CONCLUSION

On the basis of simulated events for primary protons with CORSIKA code is obtained the lateral distribution function of atmospheric Cherenkov light in extensive air showers for configuration of the Tunka-25 EAS array at the highest energies $E \geq 10^{16}$ eV. Using results of this simulation we obtained the parameters of lateral distribution function of the Cherenkov radiation as a functions of the primary energy for primary protons. The extrapolation of the Cherenkov light LDF parameterization of the obtained data with CORSIKA program at the energy range $10^{16}$- $2 \cdot 10^{18}$ eV is obtained.

The main advantage of the given approach consists of the possibility to make a library of LDF samples which could be utilized for analysis of real events which detected with the ultrahigh energy EAS arraies and reconstruction of the primary cosmic rays energy spectrum and mass composition.